\documentclass[12pt,preprint]{aastex}

\usepackage{verbatim}

\newcommand{\Ha}{H$\alpha$}
\newcommand{\ten}{$10^{10}$}
\newcommand{\erg}{ergs cm$^{-2}$ s$^{-1}$}
\newcommand{\pho}{photons s$^{-1}$ cm$^{-2}$ keV$^{-1}$}

\newcommand{\RHESSI}{$\it RHESSI$}

\shorttitle{CONTINUUM ENHANCEMENT \& HXR EMISSION}
\shortauthors{CHEN \& DING}

\begin{document}

\title{
ON THE RELATIONSHIP BETWEEN THE CONTINUUM ENHANCEMENT AND HARD X-RAY EMISSION
IN A WHITE-LIGHT FLARE}
\author{Q. R. Chen and M. D. Ding}
\affil {Department of Astronomy, Nanjing University, Nanjing 210093, China}

\begin{abstract}
We investigate the relationship between 
the continuum enhancement and the hard X-ray (HXR) emission 
of a white-light flare on 2002 September 29.
By reconstructing the \RHESSI\ HXR images 
in the impulsive phase, 
we find two bright conjugate footpoints (FPs) 
on the two sides of the magnetic neutral line.
Using the thick-target model and assuming a low-energy cutoff of 20 keV,
the energy fluxes of non-thermal electron beams bombarding FPs A and B 
are estimated to be 1.0 $\times$ \ten\ and 0.8 $\times$ \ten\ \erg, 
respectively.
However, the continuum enhancement at the two FPs 
is not simply proportional to the electron beam flux.
The continuum emission at FP B is relatively strong 
with a maximum enhancement of $\sim 8$\% 
and correlates temporally well with the HXR profile; 
however, that at FP A is less significant with an enhancement of 
only $\sim 4-5$\%,
regardless of the relatively strong beam flux.
By carefully inspecting the \Ha\ line profiles, 
we ascribe such a contrast to different atmospheric conditions at the two FPs.
The \Ha\ line profile at FP B exhibits a relatively weak amplitude
with a pronounced central reversal, while the profile at FP A is
fairly strong without a visible central reversal.
This indicates that in the early impulsive phase of the flare, 
the local atmosphere at FP A has been appreciably heated 
and the coronal pressure is high enough to prevent most high-energy electrons
from penetrating into the deeper atmosphere;
while at FP B, the atmosphere has not been fully heated,
the electron beam can effectively heat the chromosphere and
produce the observed continuum enhancement via 
the radiative backwarming effect.
\end{abstract}


\section{INTRODUCTION}
White-light flares (WLFs) are rare energetic events characterized by 
a visible continuum enhancement to a few or tens of percent, 
which imposes strict constraints on the modeling of solar flares 
in terms of energy release and transport processes in the impulsive phase.
According to the spectral features, 
two types of WLFs have been proposed \citep{mac86}
and such a category greatly facilitates our understanding of 
the physical conditions and heating mechanisms of WLFs.
The spectra of type I WLFs show a Balmer and Paschen jump,
strong and broadened hydrogen Balmer lines,
and a continuum enhancement that is well correlated with the HXR emission 
and microwave bursts \citep{fan95}.
However, type II WLFs do not show the above spectral features \citep{din99}.

The continuum enhancement in WLFs is primarily associated with 
the impulsive phase \citep{hud92,nei93a}
and often persists after the maximum phase 
\citep{hud92,mat03}.
The close temporal correlation between the continuum enhancement and 
the HXR and microwave emission in type I WLFs indicates that 
such WLFs are heated by energy deposition of non-thermal electrons 
in the chromosphere.
This process can be diagnosed using the \Ha\ line, 
which appears to be significantly enhanced and Stark broadened
\citep{can84,fan93}.
However, in most cases, direct collisional heating by the electron beam 
in the lower chromosphere and below, where the continuum emission originates,
is hardly effective 
because only electrons with very high energies can reach there
\citep{lin76,nei93b}.
Non-LTE computations also show that beam precipitation 
cannot produce the continuum enhancement directly 
\citep[e.g.,][]{liu01,din03b}.
Therefore, the continuum enhancement is supposed to be produced 
indirectly via the radiative backwarming effect 
\citep{mac89,met90a,met90b,din03b}.
This scenario assumes that non-thermal electrons, whose energies are not 
necessarily very high, heat the chromosphere first, and then 
the enhanced radiation from the upper layers is transported into 
deeper layers and causes a heating there.
On the other hand, some 
authors also attempted to investigate 
the spatial coincidence between the continuum enhancement 
and the HXR emission.
While in some WLFs, such a spatial coincidence holds well 
\citep{mat03,met03,xu04};
in some other cases, it does not 
\citep{syl00,mat02}.

In the last decade, the white-light data from the aspect camera of 
{\it YOHKOH}/SXT \citep{tsu91}
provide the first chance to study WLFs from space
(e.g., Hudson et al. 1992; Matthews et al. 2003 and references therein).
The {\it Transition Region and Coronal Explorer} ({\it TRACE})
is also capable to observe the continuum emission 
in a wide wavelength range covering the visible band \citep[see][]{met03}.
However, coincident HXR observation from 
{\it YOHKOH}/HXT \citep{kos91}
is somewhat limited by the low energy resolution 
since the HXT has only 4 broad energy bands (L, M1, M2, and H bands). 
The recently launched 
{\it Reuven Ramaty High-Energy Solar Spectroscopic Imager} (\RHESSI) 
provides unprecedented high resolution imaging spectroscopy 
\citep{lin02}.
This, together with the ground-based optical spectroscopy,
allows us to quantitatively investigate the temporal and spatial 
relationship between the continuum enhancement and non-thermal electrons 
producing the HXR emission in solar flares.

An M2.6/2B WLF on 2002 September 29 was simultaneously observed by 
the imaging spectrograph of 
the Solar Tower Telescope of Nanjing University \citep{hua95} and by \RHESSI.
A preliminary analysis of observational aspects for this flare has 
been presented in a previous paper \citep[hereafter Paper I]{din03a}.
A multi-wavelength analysis of this flare was also 
carried out by \citet{kul04}. 
In this paper, we perform a quantitative analysis of this flare by 
deriving the energy flux of 
non-thermal electrons and discussing the origin of the continuum enhancement
in terms of current WLF models. 

\section{OBSERVATIONS AND DATA ANALYSIS}
We first give a brief description of the \Ha\ and HXR emission of this flare, 
as presented in Paper I. 
This M2.6/2B flare, associated with a filament eruption, 
occurred at NOAA 0134 (N12\arcdeg, E21\arcdeg) on 2002 September 29. 
It started at 06:32 and peaked at 06:39 UT.
As in Paper I,
we pay attention to two main \Ha\ kernels, 
which are located at different magnetic polarities (see Figure 4).
In particular, we select two points (A and B) representative of the two 
kernels to check their evolutionary behaviors based on the signatures
of the \Ha\ line profile (see \S3.2).
Point A, at the center of the first kernel, is already hot 
at the start of ground-based observations and cools down gradually.
Point B, at the center of the second kernel (also the brightest kernel),
is relatively cool at first and is heated rapidly in the impulsive phase.
The continuum enhancement (calculated at \Ha+6 \AA) at Point B 
rises rapidly and reaches its maximum ($\sim 8$\%) roughly coincident with 
the peak of the 25--50 keV HXR emission.
It is interesting that the maximum continuum enhancement at Point B
is nearly twice that at Point A. 
To study the HXR emission, we first 
use the CLEAN algorithm \citep[see, e.g.,][]{kru02} to reconstruct 
the HXR images.
A strong HXR source appears to encompass both kernels
in the early impulsive phase,
and it then shows a motion across the magnetic neutral line.
Compared to data from the 
{\em Solar and Heliospheric Observatory}
MDI magnetogram, the bright HXR source
seems to straddle over
the magnetic neutral line at earlier times;
therefore, it is thought to contain two 
spatially unresolved FP sources;
the motion of the HXR source reflects a change of
the relative weights of its two components.

In addition, we employ the Maximum Entropy Method
(MEM) algorithm provided by 
the \RHESSI\ imaging software \citep{hur02}
to reconstruct HXR images around the peak of the impulsive phase. 
It is worth noting that, the CLEAN algorithm is 
a straight forward iterative algorithm involving a convolution of
source emission with instrumental Point Spread Function (PSF);
thus, it often gives diffuse images with large FWHM 
\citep[see, e.g.,][]{asc04}.
In comparison, 
the MEM algorithm \citep{sat99} generally 
yields relatively sharp images. 
In this paper, we use both the CLEAN and MEM algorithms
for different purposes.
Except for the integration time and energy band,
the imaging parameters that are explicitly set in this paper
are the same as that in Paper I for consistence. In summary, 
we use detectors 3 through 8 in image reconstruction
(thus with a spatial resolution of $\sim 7$\arcsec),
and set the image center at (--290\arcsec, 90\arcsec), 
the FOV of $64\arcsec \times 64\arcsec$, 
and the pixel size of $2\arcsec \times 2\arcsec$;
all the other parameters are taken at their default. 

Figure 1 shows the 15--50 keV HXR image in the impulsive phase 
with an integration time from 06:36:00 to 06:36:30 UT.
As expected, two conjugate HXR FPs ($black\ contours$), 
located at different magnetic polarities,
are clearly resolved by the MEM algorithm,
in comparison to the elongated CLEAN image ($grey\ scale$). 
Taking into account the spatial resolution,
the centroids of the two HXR FPs coincide well with 
Points A and B ($plus\ signs$)
in the two main \Ha\ kernels, respectively.
Moreover, a close spatial correspondence between the continuum emission 
($white\ contours$) and the HXR emission at FP B
is clearly seen from Figure 1.
Note that we draw in the figure two boxes that encompass the two HXR FPs
in order to deduce the photon spectra of them.
The result revealed in Figure 1 confirms our previous speculation
that the HXR source reconstructed with the CLEAN algorithm
is in fact two spatially unresolved FP sources (see Paper I). 

We also try other image reconstruction algorithms 
and find that the MEM images can be largely reproduced 
by the Pixon algorithm \citep{met96} that usually 
gives superior noise suppression and photometric accuracy,
but is very time-consuming.
Thus, the double-footpoint structure in the MEM images should be real,
even though the {\it RHESSI} MEM software may not ensure proper
photometric convergence especially when 
there are too many freedoms \citep{asc04}.
Further investigation on this topic is out of the scope of this paper.

We then reconstruct HXR images in 11 logarithmically spaced energy bands 
from 10 keV to $\sim 100$ keV, 
for imaging spectroscopy in the impulsive phase.
Figure 2 shows a number of selected MEM images
with pronounced features,
together with the CLEAN images for comparison.
\citet{asc04} have revealed that the CLEAN algorithm yields
a better photometric convergence than the MEM algorithm.
Therefore, we further integrate the photon fluxes over the two boxes A and B, 
respectively, using the CLEAN images rather than the MEM images. 
Figure 3 plots the photon spectra for the two FPs.

We finally reconstruct HXR images in two broad energy bands 
(12--25 keV and 25--50 keV) every 3 s with the CLEAN algorithm. 
The integration time is $\sim 4$ s.
The HXR time profiles at the two FPs are then extracted,
which are plotted in Figure 4. 

\section{RESULTS AND DISCUSSIONS}
\subsection{NON-THERMAL EMISSION IN THE IMPULSIVE PHASE} \label{bozomath}
\RHESSI\ provides for the first time high spatial and spectral resolution
imaging spectroscopy for HXR features of solar flares. 
It is seen from Figure 2 that in the impulsive phase, 
the HXR emission exhibits an evident migration from FP B to FP A 
with increasing energies.
Below $\sim 15$ keV the emission comes mainly from FP B 
while above $\sim 25$ keV FP A is dominant.
At intermediate energies the emission from the two FPs 
is of comparable magnitude.

We then fit the non-thermal component of the photon spectra 
at the two FPs, respectively.
In order to avoid possible thermal contamination,
the photon spectra are fitted above $\sim 15$ keV.
Figure 3 shows that the photon spectra at the two FPs can both be 
well fitted with a single power law.
FP A has a photon flux of 0.10 \pho\ at 50 keV and a spectral index of 
$\gamma_{A}=4.2$, while FP B has a photon flux of 0.04 \pho\ at 50 keV 
and a spectral index of $\gamma_{B}=4.7$.
Thus, the photon spectrum of FP A is slightly harder than that of FP B. 
Considering the uncertainty in defining the areas for flux integration
and spectral fitting,
such a difference is not significant for the two conjugate FPs,
which are bombarded by electron beams whose spectral indices 
are generally assumed to be approximately equal. 

Under the assumption that the non-thermal HXR emission at both FPs 
is produced via the thick-target bremsstrahlung \citep{bro71}
by electrons whose distribution is a single power law 
with a spectral index of $\delta=\gamma+1$ 
and a low-energy cutoff of 20 keV,
we first derive the total power of non-thermal electrons,
$P_{20}$ (ergs s$^{-1}$),
from the photon spectra presented above 
and then deduce the spatial distribution of energy flux, $F_{20}$ (\erg),
with the total power partitioned to each pixel
whose weight is proportional to the corresponding photon intensity.
This is formulated as 
\begin{equation}
(F_{20})_{ij}=\frac{P_{20}}{A_{ij}} \frac{I_{ij}}{\sum_{ij} {I_{ij}}},
\end{equation}
where $(F_{20})_{ij}$, $I_{ij}$, and $A_{ij}$ are 
the energy flux, photon intensity, and area at pixel ($i,j$), respectively.
Finally, we search for the maximum energy fluxes within the two FPs, 
which are found to be 1.0 $\times$ \ten\ and 0.8 $\times$ \ten\ \erg\ 
at FPs A and B, respectively.
We will show in the following that electron beams with such energy fluxes
meet well the requirement for producing the continuum enhancement
observed in this WLF.

\subsection{RELATIONSHIP BETWEEN THE CONTINUUM ENHANCEMENT
AND NON-THERMAL ELECTRONS} 
It is seen from Figure 4 that FP B exhibits a significant continuum enhancement 
in the impulsive phase that reaches a peak of $\sim 8$\% at around 06:36:35 UT.
Moreover, the temporal evolution of the continuum enhancement shows a fairly
well correlation with the 25--50 keV HXR emission.
This fact indicates that the continuum enhancement is most probably related to 
the precipitation of non-thermal electrons into the chromosphere. 
In comparison, the continuum enhancement at FP A is less significant
while the HXR emission there seems stronger than that at FP B.
To get a quantitative view between the continuum emission and non-thermal
electrons, we have further derived the energy content of the electron beams
at the two FPs (see \S3.1).
The results show that in the impulsive phase, the energy flux of 
non-thermal electrons precipitating at FP B is slightly less than that at FP A.
Therefore, there arises an interesting question:
why a stronger electron beam at FP A results in a weaker 
continuum enhancement?

To answer the question about the different responses of the continuum emission
to the non-thermal electrons at the two FPs,
we need to check carefully the \Ha\ spectral signatures that 
provide a clue to the atmospheric heating there.
Generally speaking, the \Ha\ line emission can be affected by 
three different mechanisms: beam precipitation of energetic electrons, 
thermal conduction, and enhanced coronal pressure. 
In some cases, specific heating mechanisms may be identified unambiguously from
the spectral signatures of the \Ha\ line profile (Canfield et al. 1984). 
Figure 5 plots the \Ha\ line profiles for the two FPs at 06:36:16 UT. 
The figure shows that
the \Ha\ line intensity at FP A is much stronger than that at FP B
at the start of ground-based observations, 
which means that the chromosphere at FP A has already been heated to 
a considerable extent before observations.
The continuum emission shows a different behavior: 
it increases rapidly at the relatively cool FP B in rough coincidence with
the HXR emission, while it varies slowly at the relatively hot FP A,
as shown in Figure 4.

\subsection{ORIGIN OF THE DIFFERENCE BETWEEN THE TWO FPs}
As shown in Figure 5, the \Ha\ profile at FP A 
is relatively strong and broad without a visible reversal, 
while that at FP B is relatively weak 
and shows an appreciable central reversal.
According to Canfield et al. (1984), 
only a high coronal pressure
can produce strong emission profiles without a central reversal,
which fits the situation of FP A.
Thus, the less significant continuum enhancement at FP A may result from 
a high coronal pressure which prevents most energetic electrons accelerated 
in the corona from precipitating deep into the chromosphere effectively. 
However, the \Ha\ profile at FP B is associated with a relatively 
low coronal pressure, 
which allows energetic electrons to easily penetrate into the chromosphere. 

We further estimate the coronal column density, $N$,
in the loop as follows, 
\begin{equation}
N=n \frac{L}{2}=\left(\frac{\rm EM}{AL}\right)^{1/2} \frac{L}{2},
\end{equation}
where ${\rm EM}$, $A$, and $L$ are the emission measure, 
the loop footpoint area, and the loop length, respectively,
which can be derived from 
the {\it GOES} soft X-ray fluxes and \RHESSI\ images.
The quantity of $N$ is estimated to be 
$\sim 1.0 \times 10^{20}$ cm$^{-2}$ 
in the impulsive phase.
Since FP A is much denser than FP B,
the coronal column density at FP A may be roughly equal to 
the value derived above.
The corresponding energy $E$, 
electrons of energy above which can penetrate to the chromosphere,
follows \citep{bro72,ver04} 
\begin{equation}
E=({3KN})^{1/2}=8.8 N_{19}^{1/2}\ {\rm (keV)}, 
\end{equation}
where $K=2 \pi e^4 \Lambda$ 
(with $e$ the electron charge and $\Lambda$ the Coulomb logarithm) and 
$N_{19}$ is the column density measured in $10^{19}$ cm$^{-2}$ .
Inserting the quantity $N$ derived above into Eq. (3) yields $E \simeq 27$ keV.
The consequence is that only $\sim 30$\% of the beam energy is deposited
into the chromosphere at FP A and
therefore the backwarming effect is not significant there.

In comparison, we believe that
electron heating of the chromosphere followed by the backwarming effect
results in the continuum enhancement at FP B.
Using the same method as in \citet{din03b},
we perform calculations that can predict the continuum enhancement
from a model atmosphere that is bombarded by an electron beam.
Figure 6 shows the continuum enhancement 
at $\lambda=6600$ \AA\ as a function of the beam energy flux.
It is seen that an electron beam with an energy flux of 
0.8 $\times$ \ten\ \erg\ can produce a continuum enhancement of $\sim 8$\%.
Thus, the energy flux derived for FP B seems enough to meet 
the energy requirement of the continuum enhancement. 
However, we should  mention that
the deduced energy flux suffers a great uncertainty
that arises indeed from the uncertainty of the low-energy cutoff
of the electron beam.
As shown in Figure 2, the nonthermal component of the HXR emission
in the two FPs is still visible below 20 keV;
therefore, if we select a low-energy cutoff lower than 20 keV,
say, 15 keV, the deduced beam energy flux will be 2--3 times
that if adopting the usually assumed low-energy cutoff of 20 keV.

According to the atmospheric models computed by \citet{din03b},
we obtain the temperature increase in the lower atmosphere in
response to the electron beam heating and the backwarming effect.
Then, we can estimate the timescale of the backwarming effect as
\begin{equation}
%\Delta t=\frac{3}{2} (n_{H}+n_{e})k \Delta T / |\Phi_{NT}-\Phi_{T}|,
\Delta t=\frac{3}{2} \frac{(n_{\rm H}+n_{e})k \Delta T}{|\Phi_{\rm NT}-\Phi_{T}|},
\end{equation}
where $\Phi_{\rm NT}$ and $\Phi_{T}$ are the radiative loss rates
in the two cases with and without electron beam heating, respectively,
the difference of which represents the heating rate due to the
backwarming effect. We find that the timescale varies from
$\la 1$ s near the temperature minimum region to $\sim 5$ s at
the layer of $\tau_{6600}=1$. In deeper layers, however, the timescale
becomes much longer and needs $\sim 20$ s, similar to the estimation
of \citet{hen90}.
As seen from Figure 4,
the time delay of the continuum enhancement with respect to
the 25--50 keV HXR emission is $\sim 15$ s, 
which may be explained partly by the timescale of radiative backwarming
and partly by the low temporal resolution of ground-based observations,
during which the repetition time for scanning is $\sim 10$ s.

\section{CONCLUSIONS}
In this paper, we discuss the relationship between 
the continuum enhancement and the HXR emission of the WLF on 2002 September 29 
in terms of current WLF models.
The WLF was simultaneously observed by a ground-based imaging spectrograph 
and by \RHESSI. The main results are as follows.

1. Two conjugate FPs are clearly resolved from the \RHESSI\ 
HXR images reconstructed with the MEM and Pixon algorithms, 
which are located on different sides of the magnetic neutral line. 
Around the peak of the impulsive phase, 
the energy fluxes of non-thermal electrons
bombarding FPs A and B are estimated to be 
1.0 $\times$ \ten\ and 0.8 $\times$ \ten\ \erg, respectively, 
in the framework of the thick-target model.

2. The continuum enhancement differs greatly at the two FPs.
At FP B, it increases rapidly in the impulsive phase 
reaching a maximum of $\sim 8$\%,
and correlates well with the 25--50 keV HXR emission.
While at FP A, it is less significant and varies slowly.
We show that at FP B, the derived energy flux of non-thermal electrons 
(0.8 $\times$ \ten\ \erg) can produce the observed continuum enhancement 
($\sim 8$\%) in terms of WLF models that invoke the radiative backwarming effect. 

3. The different behaviors of the continuum emission at the two FPs
can be explained by different atmospheric conditions,
which are revealed by the \Ha\ line profiles. 
The \Ha\ spectral signatures indicate that at FP A, 
the atmosphere has been heated considerably and 
the coronal pressure is high in the early impulsive phase, 
which prevents non-thermal electrons effectively penetrating 
into the chromosphere;
however, at FP B, the preflare heating is relatively low,
which allows an electron beam to 
easily penetrate into the chromosphere and
produce the observed continuum enhancement 
via the radiative backwarming effect.

\acknowledgments
We would like to thank the referee for valuable comments
that led to an improvement of the paper.
We are very grateful to the \RHESSI\ team 
for providing the observational data
and well developed analysis softwares.
This work was supported by TRAPOYT, NKBRSF under grant G20000784,
and NSFC under grants 10025315, 10221001, and 10333040,
and FANEDD under grant 200226.

\clearpage

\begin{figure}
\epsscale{.80}
\plotone{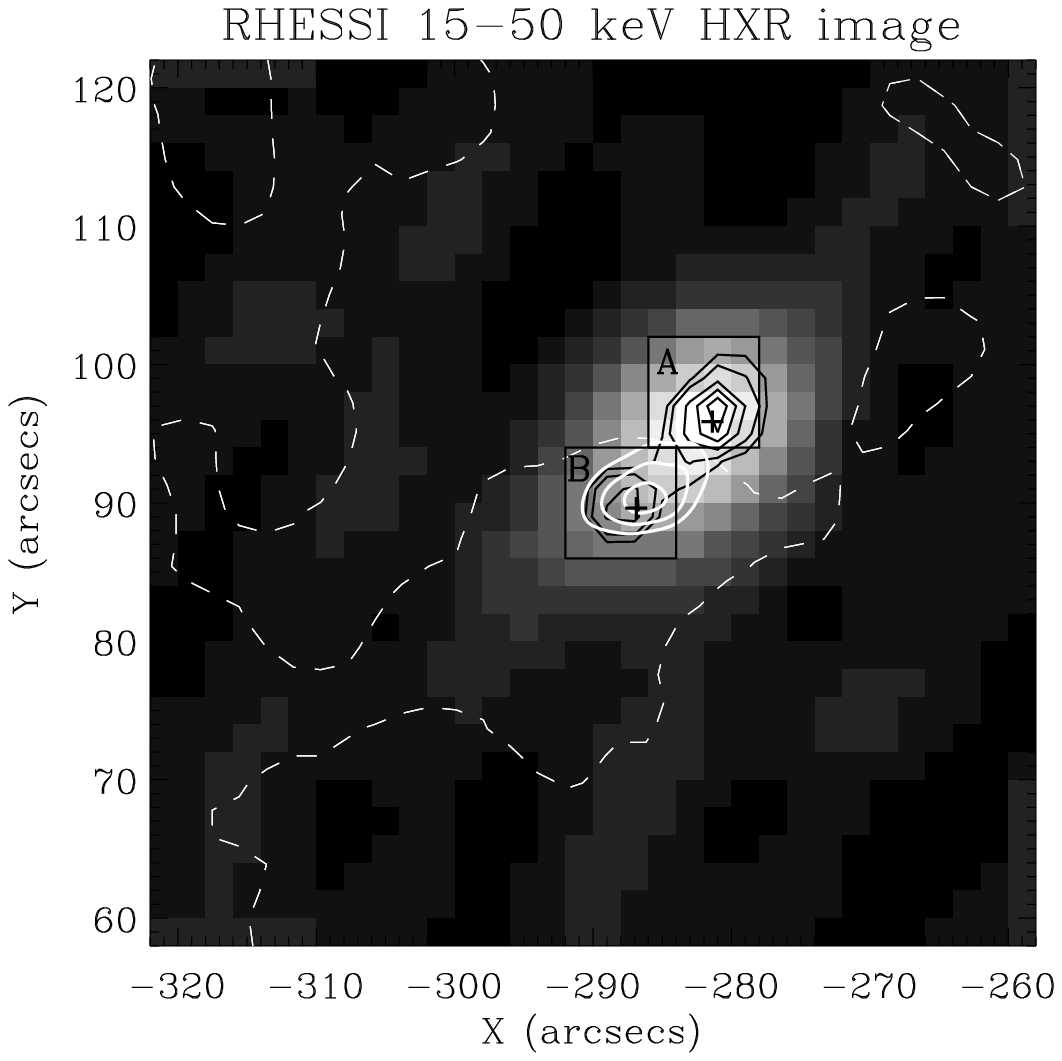}
\caption{
\RHESSI\ 15--50 keV HXR image in the impulsive phase with 
an integration time from 06:36:00 to 06:36:30 UT. The MEM image 
($black\ contours$, with levels of 10, 20, 40, 60, and 80\%)
shows two well resolved HXR FPs across the magnetic neutral line
($white\ dashed\ line$). 
The CLEAN image ($grey\ scale$) shows an elongated bright source 
covering both magnetic polarities. 
The two boxes indicate the areas, covering FPs A and B, respectively,
over which the photon fluxes in Figs. 3--4 are integrated.
Also shown in the figure is the continuum emission at 06:36:16 UT
($white\ contours$, with levels of 50, 70, and 90\%).
\label{fig1}}
\end{figure}
\clearpage

\begin{figure}
\epsscale{1.00}
\plotone{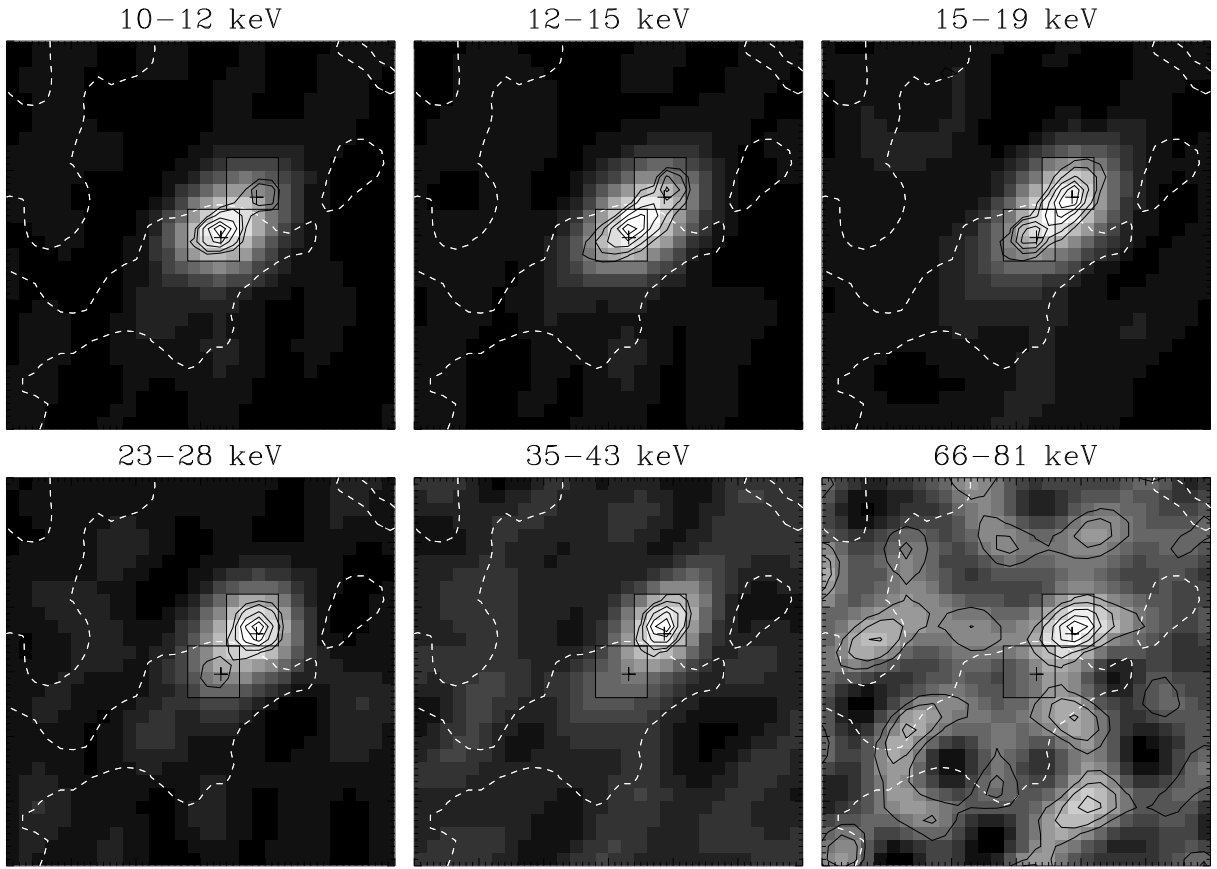}
\caption{
\RHESSI\ HXR images of the flare at different energy bands 
from 06:36:00 to 06:36:30 UT, 
reconstructed with the CLEAN algorithm ($grey\ scale$).
Superposed contours (with levels of 10, 20, 40, 60, and 80\%)
are those reconstructed with the MEM algorithm.
\label{fig2}}
\end{figure}
\clearpage

\begin{figure}
\epsscale{1.00}
\plotone{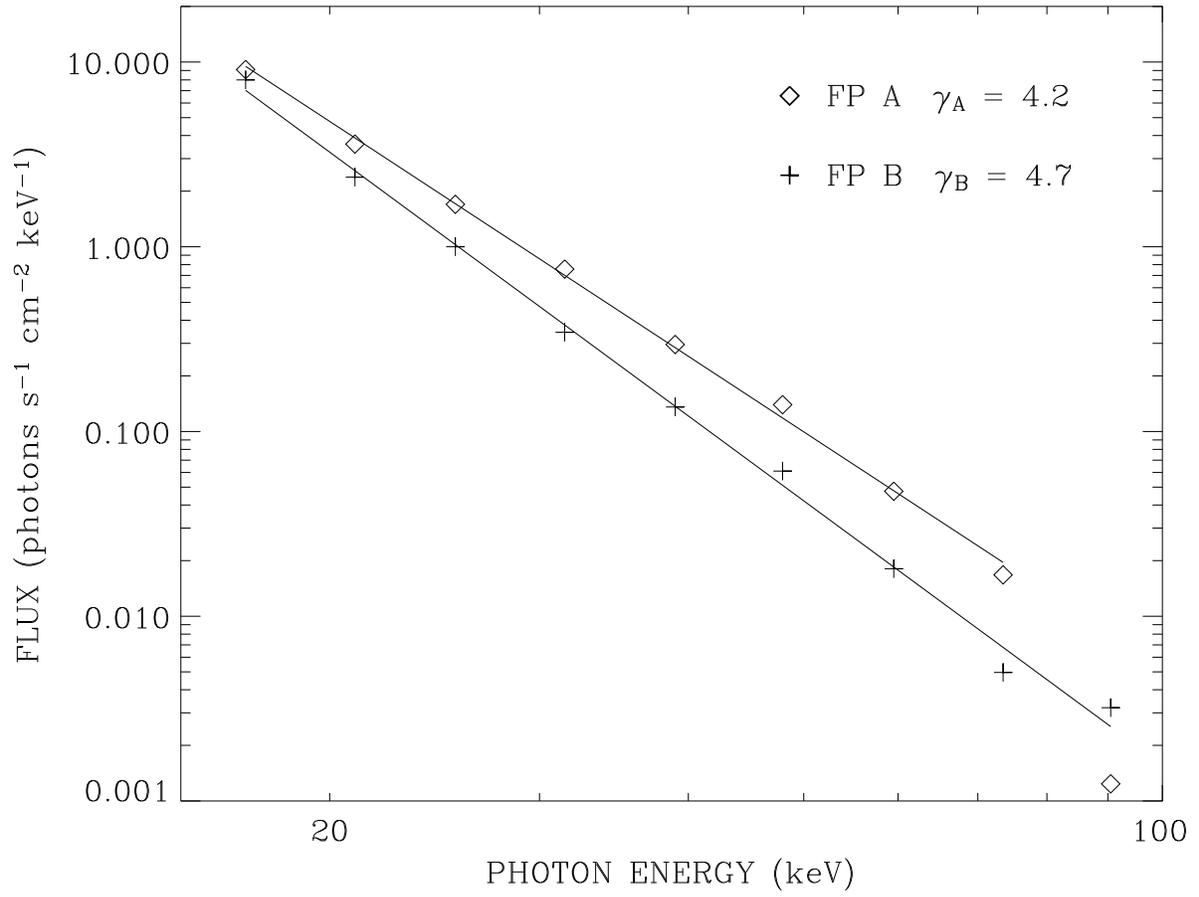}
\caption{Photon spectra for the two FPs, and their power-law fitting.
\label{fig3}}
\end{figure}
\clearpage

\begin{figure}
\epsscale{0.70}
\plotone{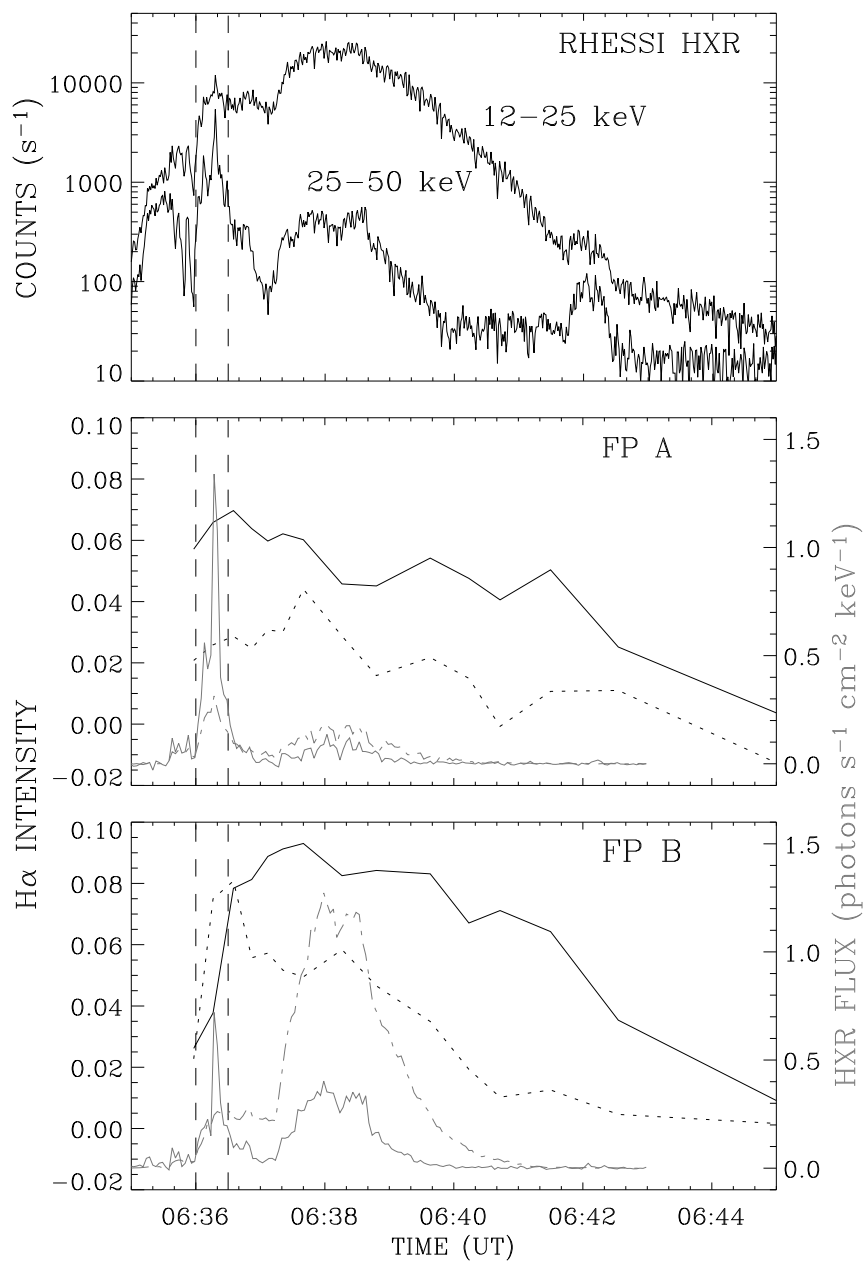}
\caption{
$Top\ panel$: 
Time profiles of the spatially integrated \RHESSI\ HXR emission 
in 12--25 and 25--50 keV energy bands. 
$Middle\ and\ bottom\ panels$:
Net increase of the emission at \Ha\ line center ($solid\ line$) and 
at \Ha+6 \AA\ (regarded as the continuum enhancement, $dotted\ line$), 
time profiles of HXR emission in 12--25 keV ($grey\ dot$-$dashed\ line$, 
scaled by 0.02) and in 25--50 keV ($grey\ solid\ line$) 
for FPs A and B, respectively. 
The two vertical bars refer to the integration time 
for HXR image reconstruction in Figs. 1--2.
\label{fig4}}
\end{figure}
\clearpage

\begin{figure}
\epsscale{.80}
\plotone{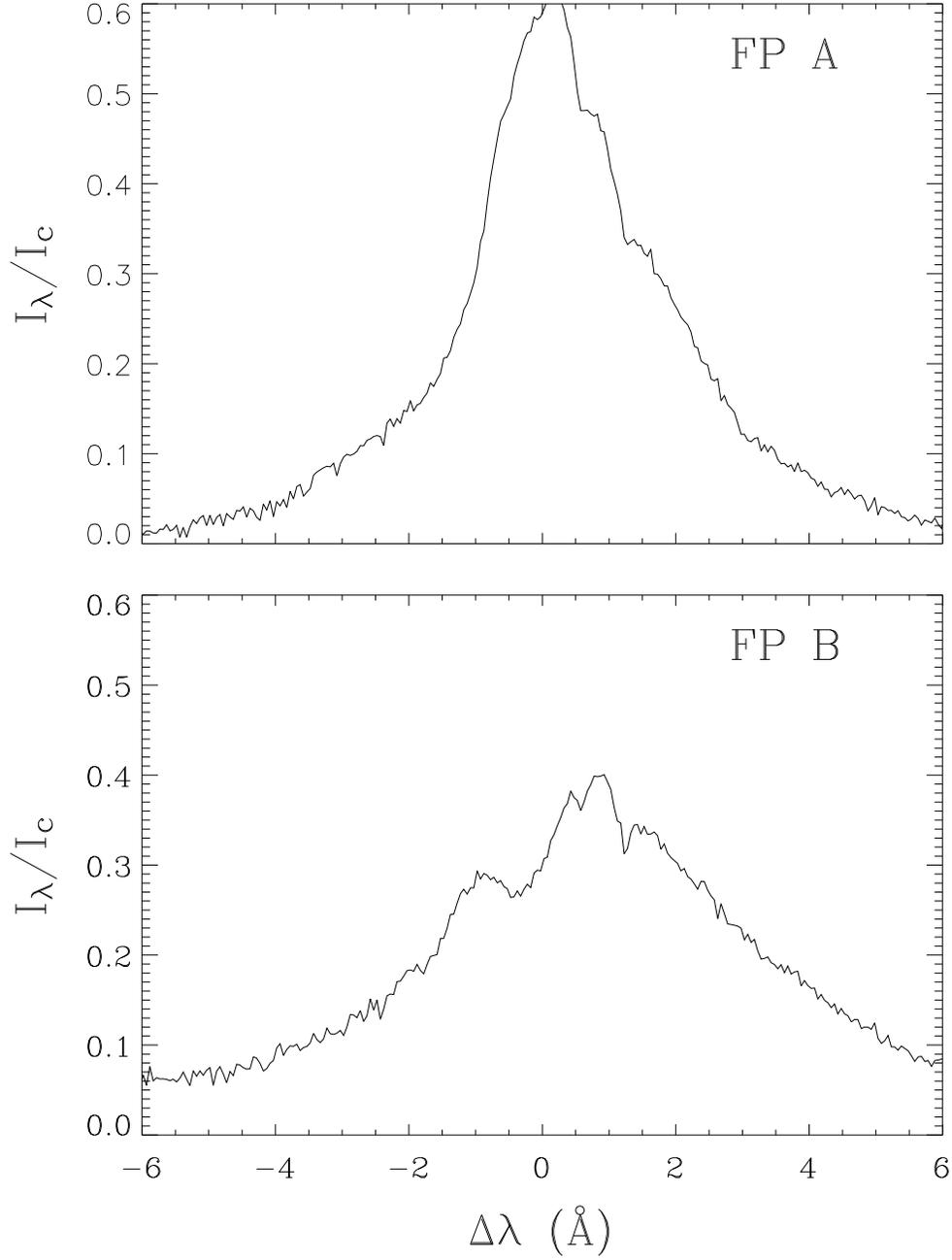}
\caption{
\Ha\ line profiles at FPs A and B at 06:36:16 UT
with the quiet-Sun profile subtracted.
The profiles are normalized by the nearby continuum.
\label{fig5}}
\end{figure} 
\clearpage

\begin{figure}
\epsscale{.80}
\plotone{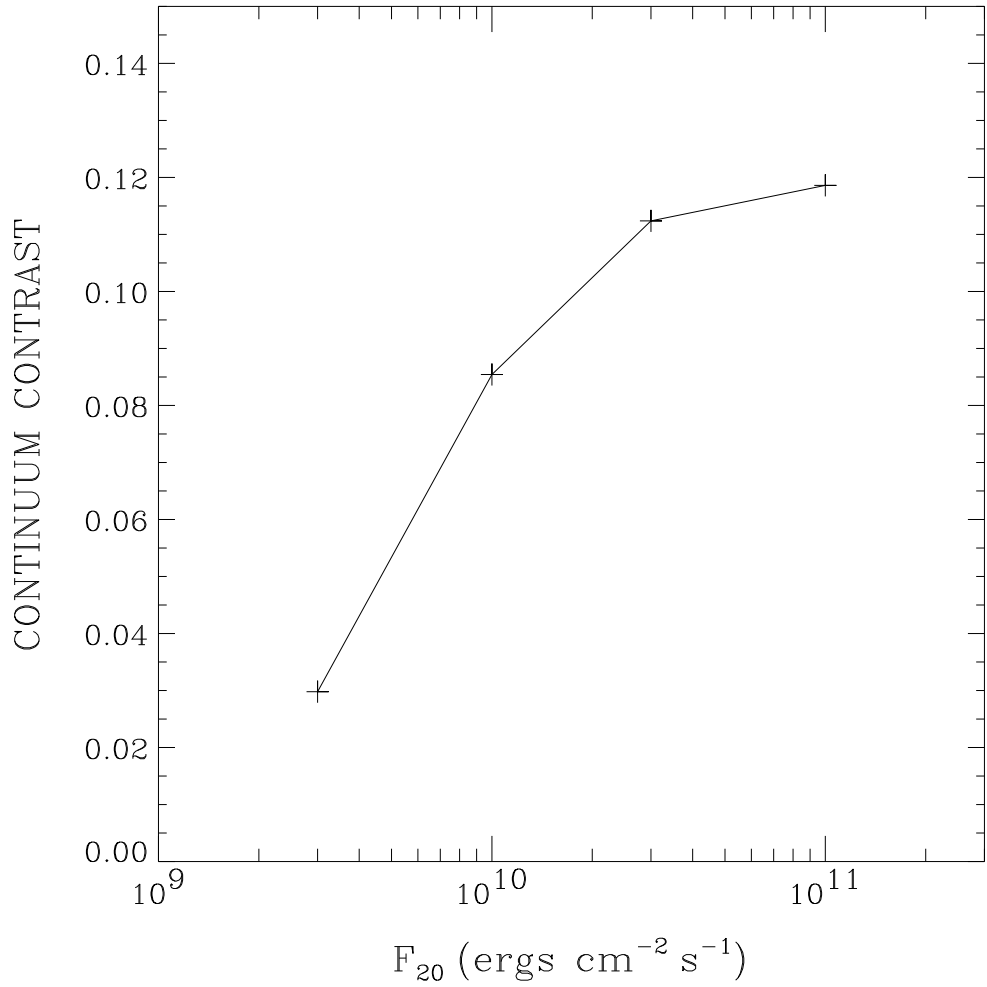}
\caption{
Theoretical prediction of the continuum enhancement at $\lambda=6600$
\AA\ as a function of the energy flux of the electron beam that
bombards the atmosphere. The calculations are similar to those of
\citet{din03b}.
\label{fig6}}
\end{figure}
\clearpage

\end{document}